\documentclass[11pt]{article}

\usepackage{times}

\newcommand{\lapp}{\mathrel{\vcenter{\hbox{\tiny \ooalign{\raise 3.25pt
        \hbox{$<$}\crcr $\sim$}}}}}
\newcommand{\gapp}{\mathrel{\vcenter{\hbox{\tiny \ooalign{\raise 3.25pt
        \hbox{$>$}\crcr $\sim$}}}}}
\newcommand{\eqdef}{\!\!\mathrel{\vcenter{\hbox{ \ooalign{\raise 4.75pt
        \hbox{${\textsf{\tiny{\,def}}}$}\crcr $=$}}}}}

\newcommand{\bi}{\begin{itemize}}
\newcommand{\ei}{\end{itemize}}

\newcommand{\forget}[1]{\iffalse#1\fi}
\newcommand{\forgetmenot}[1]{\iftrue#1\fi}

\newcommand{\be}{\begin{equation}}
\newcommand{\ee}{\end{equation}}

\renewcommand{\:}[2]{{\textstyle\frac{#1}{#2}}}
\renewcommand{\;}[2]{{\frac{#1}{#2}}}

\newcommand{\ba}{\begin{eqnarray}}
\newcommand{\ea}{\end{eqnarray}}
\newcommand{\del}{\nabla}
\newcommand{\pha}{\phantom{h}}

\renewcommand{\div}{{\mathsf{div}}\,}
\newcommand{\curl}{{\mathsf{curl}}\,}

\newcommand{\li}{&}

\newcommand{\<}{\langle}
\renewcommand{\>}{\rangle}

\newcommand{\eff}{^{^{_{_{\tiny\mathsf{eff}}}}}}

\def\half{{1\over2}}

\def\tg{\widetilde{g}}

\def\tR{\widetilde{R}}

\def\tLmatter{\widetilde{\cal L}_{\rm matter}}

\def\tT{\widetilde{T}}

\def\tnabla{\widetilde{\nabla}}

\newcommand{\udot}{\dot{u}}

\newcommand{\sdel}{\widetilde{\del}}
\newcommand{\ghat}{\hat{g}}
\newcommand{\uhat}{\hat{u}}
\newcommand{\thhat}{\hat{\theta}}
\newcommand{\rothat}{\hat{\omega}}
\newcommand{\ahat}{\hat{\dot{u}}}
\newcommand{\delhat}{\widehat{\del}}

\newcommand{\Ghat}{\hat{G}}
\newcommand{\qdot}{\dot{Q}}

\newcommand{\dd}{\hbox{d}}

\newcommand{\astroncite}[2]{\relax}

\begin{document}

\title{Inhomogeneous cosmologies, the Copernican principle and the cosmic
microwave background: More on the EGS theorem}

\author{C. A. Clarkson$^{1,2}$\footnote{Email:~\texttt{clarkson@maths.uct.ac.za}},
 A. A. Coley$^1$\footnote{Email:~\texttt{aac@mathstat.dal.ca}},
 E. S. D. O'Neill$^1$\footnote{Email:~\texttt{oneill@mathstat.dal.ca}},
  R. A. Sussman$^3$\footnote{Email:~\texttt{sussman@nuclecu.unam.mx}}
   and R. K. Barrett$^4$\footnote{Email:~\texttt{richard@astro.gla.ac.uk}}\\
\small
$^1$~Department of Mathematics and Statistics, Dalhousie University,\\
\small
 Halifax, Nova Scotia B3H 3J5, Canada.\\
\small
$^2$~Relativity and Cosmology Group, Department of Mathematics and Applied
Mathematics,\\ \small University of Cape Town,
Rondebosch 7701, Cape Town, South Africa.\\
\small
$^3$~Instituto de Ciencias Nucleares,  Apartado Postal 70543, UNAM, 
M\'exico D. F., 04510, M\'exico.\\
\small
$^4$~Astronomy and Astrophysics group, Department of Physics and
Astronomy,\\\small University of Glasgow, University Avenue, Glasgow, G12 8QQ,
UK.}

\maketitle

\begin{abstract}

We discuss inhomogeneous cosmological models which satisfy the Copernican
principle. We construct some inhomogeneous cosmological models starting from
the ansatz that the all the observers in the models view an isotropic cosmic
microwave background. We discuss  multi-fluid models, and illustrate how more
general inhomogeneous models may be derived, both in General Relativity and in
scalar-tensor theories of gravity. Thus we illustrate that the cosmological
principle, the assumption that the Universe we live in is spatially
homogeneous, does not necessarily follow from the Copernican principle and the
high isotropy of the cosmic microwave background. We also present some new
conformally flat two-fluid solutions of Einstein's field equations.

\end{abstract}


\section{Introduction}

The standard model of cosmology rests on several fundamental assumptions. As
with any theoretical model of a physical system, it is crucial that these
assumptions are identified and tested wherever possible, in order that the
proposed model be considered acceptable. The standard model of cosmology is
founded on the (perturbed) spatially homogeneous and isotropic cosmological
models of Friedman-Lema\^\i tre-Robertson-Walker (FLRW), which are derived from
the {cosmological principle}. The cosmological principle may be taken to state
that the universe is spatially homogeneous \cite{ellis75}. This is a strong
assumption; considerably stronger than the {Copernican principle} which says
that we are not at a special location in the Universe. Regardless of which
principle one cares to take, when these are combined with \emph{assumed}
perfect isotropy about ourselves, on all scales, we arrive at the FLRW models.
If we assume such perfect isotropy about us without a `mediocrity' principle,
then we must be at a center of symmetry. Obviously, the properties of a
spatially homogeneous universe can be radically different from a spherically
symmetric one, for example, and it is therefore of fundamental importance to
find some method to test the assumption of homogeneity, and to identify what
exactly happens if any of the assumptions are relaxed. This is particularly
important, since studying inhomogeneous models will allow us to identify some
possible tests of non-homogeneity in the universe.

What evidence do we have that the universe is so isotropic about us? Obviously
the strongest and most important piece of evidence for this is the extremely
high isotropy of the cosmic microwave background (CMB) which is isotropic to
one part in $10^5$. The question is: In what context can we infer spatial
homogeneity from our observations of the CMB? Without the Copernican principle
(or something similar) the answer is obviously not~-- we may be located at the
`center' of the universe and thus see the CMB isotropically distributed about
us, whereas all other observers would not have such a unique and privileged
view. However, if we assume the Copernican principle (i.e., we assume that all
or `most' observers in the universe see the CMB to be as isotropic as we see
it) can we infer homogeneity on the basis of the CMB alone?

The first attempt to answer this question resulted in a theorem by Ehlers,
Geren and Sachs \cite{EGS} (hereafter, EGS) which states that if all observers
in an expanding dust universe see an isotropic radiation field then that
spacetime is homogeneous and isotropic (and therefore FLRW). This can trivially
be generalised to the case of a geodesic and barotropic perfect
fluid~\cite{egsother}. The `isotropic radiation field' is implicitly identified
with the CMB. However, as has been emphasised
recently~\cite{cla-bar99,ferran-92}, the resulting spacetime will be FLRW only
if the matter content is of perfect fluid form, and the observers geodesic and
irrotational.  This work  has been extended \cite{cla-bar99} to include
inhomogeneous universe models with non-geodesic observers. That is,
inhomogeneous spacetimes have been found which also allow every observer to see
an isotropic CMB. It has also been shown that a significant subset of these
models are consistent with other observational constraints, regardless of
observer position \cite{BC}. This means that these models are consistent with
observations on all scales even when the Copernican principle is taken into
account~-- and yet the models are significantly inhomogeneous. However, a
problem of these models is that the non-geodesic motion of the observers
prohibits a barotropic equation of state for the matter (although the models
admit a thermodynamic scheme). There have been recent developments along this
line, where in~\cite{shearfree} a realistic multi-fluid form of the matter was
proposed (although the models used were slightly different from those
in~\cite{cla-bar99} which allow an isotropic radiation field). Alternatively, a
fractal distribution may be more appropriate~\cite{fractal}. Other recent work
concerning rotating and anisotropic cosmological models also supports these
results \cite{carn,ob1,ob2,LNW}; other non-FLRW models exist which admit an
isotropic radiation field.

The purpose of this paper is to discuss  more general cosmological models which
allow an isotropic radiation field. Specifically, we wish to discuss models
with  `realistic' matter  in order to demonstrate that there exist physically
viable inhomogeneous cosmological models which will allow an isotropic
radiation field by construction but are not FLRW. The recent supernovae data
 imply an accelerated expansion rate in the universe: within
 the standard model this implies some sort of negative pressure, be it
a cosmological constant  or quintessence or some other type of exotic matter.
Therefore, we will consider here not just traditional barotropic perfect fluid
matter, but more exotic forms, such as scalar fields and varying $\Lambda$
models.

Spacetimes which allow all observers the view of an isotropic CMB must satisfy
the `isotropic radiation field theorem'. The isotropic radiation field theorem
may be derived from the Einstein-Boltzmann equations for photons in a curved
spacetime. It is easy to show from the multipole expansions of~\cite{egsother}
that a spacetime with an isotropic radiation field must have the velocity
field, $u^a$, of the photons being shearfree and obeying
\be
\udot_a=\sdel_aQ,~~~\theta=3\dot{Q},\label{irf}
\ee
where $Q$ is a function of the energy density of the radiation field. Any
observers traveling on this congruence will observe the isotropic radiation.
This velocity field is also a conformal Killing vector of the spacetime. In
fact, a spacetime admitting an isotropic radiation field must be conformally
stationary, and we use this fact to construct some `generalised-EGS'
spacetimes.

In the following we show how some irrotational multifluid spacetimes may be
constructed, which satisfy~(\ref{irf}). To this end, we consider two
non-comoving perfect fluids, which can be interacting or non-interacting, and
may or may not admit  barotropic equations of state for the fluids. The fluids
are chosen to be non-comoving to allow for energy flux and anisotropic
pressures in the energy momentum tensor; otherwise the models may be written as
a single perfect fluid, which are the models studied in~\cite{cla-bar99,BC}. In
a similar vein we consider models with a perfect fluid and scalar field; in
contrast to usual work on mixtures of this kind, we allow for the case where
the  scalar field has a spatial gradient relative to the perfect fluid, which
we take to be comoving with the isotropic radiation (test) field. In both cases
we take one of the perfect fluids to be `comoving' with the radiation field;
that is one of the fluid velocities will be (parallel to) the timelike
conformal Killing vector of the spacetime. We then consider some models with
non-zero heat flux, but zero anisotropic pressure, which were previously
considered in \cite{CJP}, which may be interpreted as `quintessence' models
with varying $\Lambda$ and energy flux.

The case of non-zero rotation has also been considered elsewhere, and simple
expanding and rotating spacetimes with plausible matter in which the observers
could measure an isotropic CMB have been constructed
\cite{ob1,ob2,carn,obukov}. This may be considered as a counter-example to
numerous claims that the rotation of the universe may be constrained by
observations of the CMB alone; such results make additional assumptions of the
matter present and the velocity field we follow in the universe.


The upshot of all this is to emphasise that the high isotropy of the CMB when
combined with the Copernican principle is simply not enough to draw conclusions
about the spatial homogeneity of our universe.

\section{Spacetimes admitting an isotropic radiation field}

We are interested in spacetimes in which the high isotropy of the CMB is
permissible for every observer. In the particular case where we have a model in
which all observers on some congruence $u^a$ see an exactly isotropic radiation
field, then this velocity field has two important properties:
\be
\del_{[a}\left(\udot_{b]}-\:13\theta
u_{b]}\right)=0=\sigma_{ab}.\label{irf2}
\ee
Writing $\udot_{a}-\:13\theta u_{a}=\del_a Q$ we see that the first condition
is equivalent to (\ref{irf}). Spacetimes admitting an isotropic radiation field
are conformally stationary, with the velocity fields of the two (conformally
related) spacetimes parallel~-- see the appendix. Now, if we were to assume
that these observers measured only dust, then that spacetime must be FLRW~--
the origional EGS theorem~\cite{EGS}.

In this paper, for simplicity, we also restrict our attention to the
irrotational case. This may be justified by the following considerations. If
part of the matter consists of a conserved comoving barotropic perfect fluid
other than radiation, or for geodesic motion with any matter source, it follows
from~(\ref{irf}) that the expansion or the rotation must be zero. For a
conserved barotropic perfect fluid, we have $\dot{u}_a=\widetilde{\nabla}_a
\phi$, and $p'\theta=\dot\phi$, where $\phi\equiv-\int dp/(\mu(p)+p)$, and
$p'=dp/d\mu$; so, $\eta_{abc}\widetilde{\nabla}^b\widetilde{\nabla}^c(Q-\phi)=
2(\frac{1}{3}-p')\theta\omega_a=0$. For geodesic motion,
$\eta_{abc}\widetilde{\nabla}^b\widetilde{\nabla}^c
Q=\frac{2}{3}\theta\omega_a=0$. However, rotating universes which allow an
isotropic radiation field have been found and discussed in some detail~--
see~\cite{ob1,ob2,obukov,carn}.

In this case the metric can then take the form
\be
ds^2=e^{2Q(t,x^\alpha)}\left\{-dt^2+H_{\alpha\beta}dx^\alpha
dx^\beta\right\}\label{metrica}
\ee
where $H_{\alpha\beta}(x^\gamma)$ can be diagonalised. If $Q=Q(t)$ then the
acceleration is zero and we recover the models studied by Coley and
MacManus\cite{CM}; indeed, even in this case (i.e., the acceleration-free case)
it follows that there are physically viable spacetimes that are not FLRW.

In order to find irrotational spacetimes with an isotropic CMB, we can simply
compute the Einstein tensor of~(\ref{metrica}), and equate with the matter we
desire. In the appendix, we discuss this computation further.

\subsection{The energy momentum tensor of multiple fluids}

In general, the energy momentum tensor,~$T_{ab}$ for any spacetime may be
decomposed with respect to the velocity field $u^a$ in the following covariant
manner:
\be
T_{ab}=\bar\mu u_au_b+\bar ph_{ab}+2\bar
q_{(a}u_{b)}+\bar\pi_{ab}.\label{Tab-general}
\ee
This decomposition allows us to make the physical interpretations that
$\bar\mu=u^au^bT_{ab}$ is the energy density, $\bar p = \:13 h^{ab}T_{ab}$ the
isotropic pressure, $\bar q_a=-h_{a}^{~b}u^cT_{bc}$ the energy or heat flux and
$\bar\pi_{ab}=T_{\<ab\>}$ the anisotropic pressure or stress. All these
quantities are interpreted by an observer traveling on the $u^a$ congruence.

\subsubsection{perfect fluids}

Consider the energy-momentum tensor due to two non-comoving perfect fluids;
\be
T_{ab}=\mu_1 u_au_b+p_1 h_{ab} +\mu_2 \tilde{u}_a\tilde{u}_b+p_2
\tilde{h}_{ab},
\ee
where $\mu_i$ are the energy densities of the fluids in each comoving frame,
and the $p_i$'s are their respective pressures. The velocity field of the
second congruence may be written as a Lorentz boost of the first;
\be
\tilde{u}^a=\gamma (u^a+v^a),
~~~\gamma=\;{1}{\sqrt{1-v^av_a}},~~~v^au_a=0.
\ee
If we write this as one fluid with respect to the $u^a$ congruence, then
$T_{ab}$ has the form of~(\ref{Tab-general}), with components
\ba
\bar\mu=u^au^bT_{ab}&=&\mu_1+\mu_2+\gamma^2v^2\left(\mu_2+
p_2\right),\nonumber\\
\bar p =\:13h^{ab}T_{ab}&=&
p_1+p_2+\:13\gamma^2v^2\left(\mu_2+p_2\right),\nonumber\\
\bar q_a=-h_{a}^{~b}u^cT_{bc}&=&\gamma^2(\mu_2+p_2)v_a,\nonumber\\
\bar\pi_{ab}=T_{\<ab\>}&=&\gamma^2(\mu_2+p_2)v_{\<a}v_{b\>}.\label{2fluid_compts}
\ea
Thus we see that the first fluid will experience an energy flux due to the
second fluid passing through their frame (provided $\mu_2+p_2\neq0$).

\subsubsection{perfect fluid plus scalar field}\label{SF}

In general the energy-momentum tensor of a scalar field~$\phi$ may be written
\be
T^{\phi}_{ab}=\phi_{,a}\phi_{,b}-g_{ab}\left(\:12\phi_{,c}\phi^{,c}+V(\phi)\right),
\ee
which, when a velocity field is specified, takes the form
\be
T^{\phi}_{ab}=\dot\phi^2u_au_b+\sdel_a\phi\sdel_b\phi -2\dot\phi
u_{(a}\sdel_{b)}\phi
-g_{ab}\left(\:12\sdel_c\phi\sdel^c\phi-\:12\dot\phi^2+V(\phi)\right).
\ee
Thus, if we add to this scalar field a perfect fluid (perfect with respect to
this $u^a$ congruence), then we find that the total or mean matter variables
become
\ba
\bar\mu&=&\mu+\mu_\phi=\mu+\:12\dot\phi^2+\:12\sdel_c\phi\sdel^c\phi
+V(\phi),
\nonumber\\
\bar
p&=&p+p_\phi=p+\:12\dot\phi^2-\:16\sdel_c\phi\sdel^c\phi-V(\phi),\nonumber\\
\bar q_a&=&q^{\phi}_{a}=-\dot\phi\sdel_a\phi,\nonumber\\
\bar\pi_{ab}&=&\pi^{\phi}_{ab}=\sdel_{\<a}\phi\sdel_{b\>}\phi.
\ea
Note that if $\sdel_a\phi=0$ then formally the total fluid takes the form of a
single perfect fluid.

We may demand that $\phi$ satisfies the Klein-Gordon equation, which may be
derived from the energy conservation equation for the scalar field,
$\del^aT^{\phi}_{ab}=0$,
\be
\;{\partial V(\phi)}{\partial\phi}=\del_a\del^a\phi=\udot^a\sdel_a\phi
+\sdel^a\sdel_a\phi-\ddot\phi-\theta\dot\phi.
\ee
However, if there is an interaction between the scalar field and some other
matter, then this equation may not hold; for example, we could have a scalar
field decaying into physical matter   in which case the energy of the scalar
field will not be conserved independently.

\subsection{Scalar Tensor Theories of Gravity}

Scalar-tensor theories, in which a long-range scalar field  combined with a
tensor field  mediate the gravitational interaction, are standard alternatives
to general relativity. The original motivation for these theories was to
incorporate a varying gravitational constant into GR  to account for alleged
discrepancies between observations and weak-field predictions of GR. A special
case of the scalar tensor theories, known as the Brans-Dicke theory of gravity
(BDT \cite{BransDicke}) (with a constant ${\omega}_{0}$ parameter), was the
original of these theories. Scalar-tensor theories occur as the low-energy
limit in supergravity theories from string theory \cite{Green} and other
higher-dimensional gravity theories \cite{Applequist}.

Recently the recovery of the EGS theorem in scalar tensor theories was
given~\cite{CCO}; geodesic observers in a scalar tensor theory of gravity
observing isotropic radiation must be in a FLRW universe. We mention here that
inhomogeneous spacetimes are possible however if the geodesic assumption is
dropped. The field equations, obtained by varying the BD action with respect to
the metric and the field $\phi$, are
\ba
G_{ab}&=&\;{8\pi}{\phi}T_{ab}+\;{\omega}{\phi^2}\left(\del_a\phi\del_b\phi-
\:12g_{ab}\del_c\phi\del^c\phi\right)+\phi^{-1}(\del_a\del_b\phi-
g_{ab}\del_c\del^c\phi)-\:12g_{ab}U(\phi).\label{einstein}
\\
&&(3+2\omega) \del_a\del^a\phi  =  8\pi T - \del_a \omega\del^a\phi+
{{dU}\over{d\phi}}
 \label{eEOMphi}
\ea
where the energy-momentum tensor of the matter, $T^{ab}$, may take any of the
usual desired forms.  In the scalar-tensor gravity theories the principle of
equivalence is guaranteed by requiring that all matter fields are minimally
coupled to the metric $g_{ab}$. Thus energy-momentum is conserved:
\be
\nabla^{a} T_{ab} = 0.\label{cons}
\ee
It is known that scalar-tensor theories can be rewritten in the conformally
related `Einstein' frame \cite{MimosoWands}, so that the models are formally
equivalent to GR coupled to a scalar field. Therefore, scalar-tensor theories
may be incorporated here as a special case of the scalar field in GR~-- see
Sec.~\ref{SF}.

\section{Some solutions}

Rather than provide an exhaustive study of multi-fluid solutions, we will
present some example of how such solutions may be derived. This is in keeping
with our aim of illustrating the existence of `realistic' inhomogeneous
cosmological solutions which satisfy the Copernican principle.

\subsection{Two perfect fluids}

For simplicity, we restrict ourselves to the case where $Q(t,x^\alpha)=Q(t,r)$
in comoving polar coordinates and we also restrict ourselves to the spherically
symmetric case, $H_{rr}=H_{\theta\theta}=H_{\phi\phi}=\exp{2B(r,\theta,\phi)}$.
In this case the  energy flux relative to the comoving $u^a$ frame is
\be
\bar
q_r=2e^{-Q}\left(Q_{,tr}-Q_{,t}Q_{,r}\right)=\gamma^2(\mu_2+p_2)v_r,~~~
\bar q_\theta=\bar q_\phi=0 \Rightarrow v_\theta=v_\phi=0.\label{qpf}
\ee
Because $v^a$ has one non-vanishing component, this implies
from~(\ref{2fluid_compts}) that ${\bar\pi}_{ab}$ must be diagonal. Calculating
the components of $\pi_{ab}$, we find that $B$ can only be a function of $r$
alone:
\ba
{\bar\pi}_{rr}&=&\;23\left[-(B''+2Q'')+(B'+2Q')^2-2Q'^2+\;1r(B'+2Q')\right]
\nonumber\\
 &=& \;23\gamma^2(\mu_2+p_2)v_r^2.\label{pipf}
\ea
Hence we find
\be
v_r=\;32\;{{\bar\pi}_{rr}}{\bar q_r}.
\ee
We may also calculate the mean energy density and pressure;
\ba
\bar\mu&=&3e^{-2Q}Q_{,t}^2-e^{-2(Q+B)}\left[2(B''+Q'')+(B'+Q')^2+
\;4r(B'+Q')\right],
\nonumber\\
&=&\mu_1+\mu_2+\gamma^2v^2\left(\mu_2+ p_2\right),\nonumber\\
\bar
p&=&3e^{2Q}\left[Q_{,t}^{2}+2Q_{,tt}\right]-e^{-2(Q+B)}\left[2(B''+Q'')+
 B'^2+5Q'^2+4Q'B'+\;4r(B'+2Q')\right]\nonumber\\
 &=& p_1+p_2+\:13\gamma^2v^2\left(\mu_2+p_2\right).\label{matter}
\ea

As yet we have only determined $v$ (and~$\gamma$), and we have three equations
relating four functions,~$\mu_i,p_i$ (Eqs. ~(\ref{matter}), and the remaining
freedom from $\bar\pi_{rr}$ and $q_r$). In principle we have the freedom to
specify one more equation relating the four free functions.
 The most obvious restrictions are barotropic equations of state for
the two fluids, $p_i=p_i(\mu_i)$, or separate energy conservation for the two
fluids, $\del^aT^i_{ab}=0$. However, if $Q$ and $B$ are specified then only one
of these types of conditions may be used in general. We can use the freedom in
the two metric functions to allow us to use both conditions if we choose. For
simplicity, we consider the case of the two fluids obeying linear equations of
state, $p_i=w_i\mu_i$.

\subsubsection{Example: $p_i=w_i\mu_i$, with $B=0$}

We have four free functions, $Q~\mu_i$, and $v_r$ together with four equations;
specifying two equations of state is then sufficient to close the system. We
may solve Eqs.~(\ref{qpf}) and~(\ref{pipf}) for $\mu_2$ and $v_r$ as functions
of $Q$ (and derivatives of). Substituting these into~(\ref{matter}) we get two
equations for $\mu_1$; requiring equality leads to a horrendous equation for
$Q$;
\ba
0&=&  ((2\,{r}^{2}+2\,w_{{1}}{r}^{2} ){Q_{{{\it rr}}} }^{2}+ (
(-w_{{1}}{r}^{2}-{r}^{2}+3\,w_{{2}}w_{{1}}{r}^{2 }+3\,w_{{2}}{r}^{2}
){Q_{{r}}}^{2}+ (6\,w_{{2}}w_{{1}}r+ 2\,w_{{1}}r \nonumber \\
&+&  4\,w_{{2}}r )Q_{{r}} + (-2\,w_{{2}}{r}^{2}-2
\,{r}^{2} )Q_{{{\it tt}}}
+ (-w_{{2}}{r}^{2}-3\,w_{{2}}w_{ {1}}{r}^{2}-{r}^{2}-3\,w_{{1}}{r}^{2}
){Q_{{t}}}^{2} ) Q_{{{\it rr}}}
\nonumber\\
&+& (-3\,w_{{2}}{r}^{2}-w_{{1}}{r}^{2} - 3\,w_{{2}}w_ {{1}}{r}^{2} - {r}^{2}
){Q_{{r}}}^{4} + (-7\,w_{{2}}r-5\,w _{{1}}r -3\,r-9\,w_{{2}} w_{1}r
){Q_{{r}}}^{3}\nonumber \\
&+&   ( (2\,w_{{2}}{r}^{2}+2\,{r}^{2} )Q_{{{\it tt}}} +  (3\,w_{{2}
}{r}^{2}+{r}^{2} +  w_{{1}}{r}^{2} +  3\,w_{{2}}w_{{1}}{r}^{2} ){Q _{{t}}}^{2}
- 4\,w_{{2}}-4\,w_{{1}}-2-6\,w_{{2}}w_{{1}} ){Q_{{r} }}^{2} \nonumber\\
&+& ( (2\,w_{{2}}r+2\,r )Q_{{{\it tt}}} +   (3
\,w_{{1}}r+w_{{2}}r+3\,w_{{2}}w_{{1}}r
+ r ){Q_{{t}}}^{2}
 +
 (-4\,w_{{2}}{r}^{2}Q_{{{\it tr}}}+4\,w_{{1}}{r}^{2}Q_{{{\it tr
}}} )Q_{{t}} )Q_{{r}}\nonumber \\
&+& 2\,w_{{2}}{r}^{2}{Q_{{{\it tr}}}} ^{2}-2\,w_{{1}}{r}^{2}{Q_{{{\it
tr}}}}^{2})/ rw_{{1}} (-Q_{{r}} -r{Q_{{r}}}^{2} +  rQ_{{{\it rr}}} )
  (1+w_2 ) .
\ea
A solution of the form $Q=a\ln t+b\ln r$ exists, provided we choose
\ba
a&=&-2\,{\frac {1+{w_{{2}}}^{2}+2\,w_{{2}}}{-8\,w_{{2}}w_{{1
}}-2\,w_{{2}}-3\,w_{{1}}+3\,{w_{{2}}}^{2}-2\,{w_{{1}}}^{2}+3\,{w_{{2}}
}^{2}w_{{1}}-6\,w_{{2}}{w_{{1}}}^{2}-1}},\\b&=&-2\,{\frac {3\,w_{{2}}w_{{1
}}+2\,w_{{2}}+w_{{1}}}{1+3\,w_{{2}}+w_{{1}}+3\,w_{{2}}w_{{1}}}}.
\ea
The only other physical constraint is that $v^2<1$, for all $t$ and $r$. In the
case where the first fluid is dust, $w_1=0$, this requires that
$-\:13<w_2<-\:16$. This also ensures that the fluids become comoving at late
times.

Thus we have demonstrated the existence of inhomogeneous two barotropic fluid
solutions of the field equations which allow the existence of isotropic
radiation. There are clearly much more general solutions than we have presented
here, our solution being a very special case.

\subsection{Perfect fluid plus scalar field revisited.}

As before, we restrict ourselves to the spherically symmetric case where $Q(t,
x^{\alpha}) = Q(t,r)$ and $H_{rr} = H_{{\theta}{\theta}} = H_{{\phi}{\phi}} =
\exp(2B(r,{\theta},{\phi}))$. In this case the energy flux relative to the
comoving $u^{a}$ frame is
\ba
{\bar q}_{r} = 2{e}^{-Q}({Q}_{,tr} - {Q}_{t}{Q}_{r}) = -{\dot
{\phi}}\sdel_{r}{\phi}, ~~~ {\bar q}_{\theta} = {\bar q}_{\phi} = 0
\Rightarrow
{\phi}(t,r).
\ea
Because ${\phi}$ is only a function of time and the spatial coordinate $r$,
this implies that ${\bar {\pi}}_{ab}$ must be diagonal. Calculating the
components of ${\bar {\pi}}_{ab}$, we find that $B$ can only be a function of r
alone. We find that
\ba
{{2}\over{3}}(\sdel_{r}\phi)^2 = {\bar {\pi}}_{rr} &=& {{2}\over {3}}[-({B}{''}
+ 2{Q}{''}) + 2{Q'}^{2} + {B'}^{2} + 4{B'}{Q'} + {{1}\over {r}}(2{Q'} + {B'})].
\ea
Hence we find
\ba
{\dot {\phi}}^{2} = {{2}\over{3}}{{{{\bar q}_{r}}^{2}} \over {{\bar
{\pi}}_{rr}}}.
\nonumber
\ea
We may also calculate the mean energy density and pressure;
\ba
{\bar {\mu}}& =& 3e^{-2Q}{\ddot Q}^{2} - e^{-2(Q + B)}[2({B}^{''} + {Q}^{''}) +
({B}' + {Q}')^{2} + {{4}\over {r}}({B}' + {Q}')]
\nonumber\\
&=& {\mu} + {{1}\over {2}}{\dot {\phi}}^{2} + {{1}\over
{2}}\sdel_{r}{\phi}\sdel^{r}{\phi} + V({\phi}),
\ea
\ba
{\bar {p}}& =& -{{1}\over {3}}[3e^{-2Q}[{\dot Q}^{2} + 2{{\ddot Q}}] - e^{-2(Q
+ B)}[4Q^{''} + 2B^{''} + 4B'Q' + 5{Q'}^{2} + {B'}^{2} + {{4}\over{r}}(B' +
2Q')]]
\nonumber\\
&=& p + {{1}\over {2}}{\dot {\phi}}^{2} - {{1}\over
{6}}\sdel_{r}{\phi}\sdel^{r}{\phi} - V({\phi}),\label{equation36}
\ea
We may impose the restriction of a barotropic equation of state for the perfect
fluid $p({\mu})$ and an energy conservation law for the fluid $\del^{a}T_{ab} =
0$. We choose the equation of state $p = w{\mu}.$ The energy conservation
equations for the perfect fluid read
\be
{\mu}_{,t} + 3{Q}_{,t}(1 + w){\mu} = 0 \label{equation37}
\ee
\be
w{\mu}' + Q'(1 + w){\mu} = 0, \label{equation38}
\ee
which when integrated imply that $Q$ must have the form
\ba
Q(t,r) = {\alpha}(t) + {\beta}(r). \label{equation39}
\ea
It follows that a specific form for the comoving energy will be
\ba
{\mu} = \exp[-(1 + w)(3{\alpha} + {\beta}/w)]. \label{equation40}
\ea
To simplify things we will assume that $B=0$, so that the metric now becomes
\ba
g_{ab} = e^{2{\alpha}(t) + 2{\beta}(r)}{\eta}_{ab}; \label{equation41}
\ea
which implies that ${\theta} = 3{\alpha}_{,t}e^{-Q}$,~${\dot u_{r}} =
{\beta}'$, and
\ba
{\bar {\pi}}_{rr}& = &{{2}\over{3}}\left[-2{\beta}{''} + 2{{\beta}'}^{2} +
{{2}\over {r}}{\beta}'\right] \label{equation42}\\ {\bar q}_{r} &=&
-2e^{-Q}{{\alpha}_{,t}}{\beta}'. \label{equation43}
\ea
Hence we find that
\be
{\dot {\phi}}^{2}  =   {{2}\over{3}} {{{{\bar q}_{r}}^{2}}\over {{\bar
{\pi}}_{rr}}}
 =   {{2}\over
{9}}{\theta}^{2}{{\beta}'}^{2}{{1}\over {[{{\beta}'}^{2} + {{{\beta}'}\over
{r}} - {{\beta}^{''}}]}}. \label{equation44}
\ee
The scalar field wave equation is
\ba
 -{{\ddot {\phi}}} + {\phi}^{''}e^{-2B} - 2{\dot Q}{\dot {\phi}}
 + (2{Q}' + {B}'){\phi}'e^{-2B} + {{2{\phi}'}\over {r}}e^{-2B}
= {{dV}\over {d{\phi}}}e^{2Q}. \label{equation45}
\ea

We will assume a solution of the form
\ba
{\phi} = {\Phi}(t) + {\Psi}(r) \label{equation46} ,
\ea
which allows the derivation of the following two equations from the scalar
field equation  when $V=0$:
\ba
{\ddot{\Phi}} + 2{\dot {\alpha}}{\dot{\Phi}} = C \label{equation47}
\ea
\ba
{\Psi}{''} + 2({\beta}' + {{1}\over {r}}){\Psi}' = C,
\label{equation48}
\ea
where C is a constant. We can rewrite the equations for heat conduction and
anisotropic pressure  as
\ba
{\dot{\Phi}}{\Psi}' = 2e^{-Q}{\dot{\alpha}}{\beta}'
\ea
\ba
{{\Psi}'}^{2} = -2{\beta}^{''} + 2{{\beta}'}^{2} + {{2}\over {r}}{\beta}'
\ea
Hence we compute the following differential equations for ${\alpha}$ and
${\beta}$:
\ba
{\ddot{\alpha}} + {\dot{\alpha}}^{2} = Ae^{\alpha}
\ea
\ba
{\beta}^{''} + {{\beta}'}^{2} + 2{{{\beta}'}\over {r}} = Be^{\beta}
\ea
These equations have non-trivial solutions implying a non-FLRW cosmology (the
FLRW limit is recovered when $\beta=0$). Hence we have shown that spacetimes
with a barotropic perfect fluid and a \emph{non-comoving} scalar field exist
which allow an isotropic radiation field for all observers, which are non-FLRW.
Clearly there are a huge number of solutions meeting this criteria; we have
demonstrated existence in this simplest of cases. These new solutions could
play an important role in cosmology, for example as a new generalisation of
quintessence.

\section{Inhomogeneous quintessential cosmologies}

Simple multifluid models can be constructed by introducing, together with a
barotropic fluid, a varying $\Lambda (t)$ term as the second fluid:

\be\rho=\mu+\Lambda,\qquad p=(\gamma-1)\mu-\Lambda \label{eqst}\ee

\noindent
where $\rho,\,p$ are ``total'' energy density and pressure obtained from the
field equations. The varying $\Lambda$ term can be interpreted as the
asymptotic state of a scalar field associated with a quintessence dominated
scenario, coexisting with a material fluid described by $\mu$. We show in this
section that such an interpretation is compatible with the asymptotic
properties of a class of simple models that allow an isotropic radiation field.

The simplest models satisfying the EGS criterion and compatible with the
decomposition (\ref{eqst}) are characterized by the conformally FLRW metric (a
particular case of~(\ref{metrica})) given by
\be\dd s^2=\frac{-\dd t^2+\dd r^2+
r^2\left(\dd\theta^2+\sin^2\theta\,\dd\varphi^2\right)}{\Phi^2},\qquad
\Phi\equiv a(t)+b(t)r^2
\label{metric}\ee
whose source is the imperfect fluid
\be T_{ab}=\rho\,u_au_b+p\,h_{ab}+2q_{(a}u_{b)}\label{source}\ee
where $u^a=\Phi\delta^a\,_t $ and $q_a=q_r\,\delta^r\,_a$, while $\rho$ and $p$
must comply with (\ref{eqst}). Applying the field equations for (\ref{metric})
and (\ref{source}) leads to
\be
8\pi\,\gamma\mu=2\left(a_{,tt}+2b+b_{,tt}r^2\right)(a+br^2)\label{eqmu}\ee
\be 8\pi \,q_r=-4b_{,t}r,\qquad 8\pi\,
q=8\pi\,|g^{ab}q_aq_b|^{1/2}=4|b_{,t}|r(a+br^2) \label{eqq}\ee
\be 8\pi\,\gamma\Lambda =L_4(t)\,r^4+2\,L_2(t)\,r^2+L_0(t)\label{eqL}\ee
where
\be L_4(t)\equiv 3b_{,t}^2\gamma-2b_{,tt}b, \quad
L_2(t)\equiv3\gamma\,a_{,t}b_{,t} -2b^2-a_{,tt}b-b_{,tt}a,  \quad
 L_0(t)\equiv
3\gamma\left(a_{,t}^2 +4ab\right)-2a\left(2b+a_{,tt}\right)
\label{eqL4}\ee
If we demand that $\Lambda$ be only a time-dependent function, we obtain
\be L_2(t)=0, \qquad
L_4(t)=0,  \qquad
  8\pi\,\Lambda=L_0(t)/\gamma \label{eqsL24}\ee
which yields differential equations that determine $a,\,b$ for a given
$\gamma(t)$, and the definition of $\Lambda(t)$.

Since we are interested in an asymptotic regime that assumes a slowly varying
$\gamma(t)$, we shall consider a constant $\gamma$. The general solution of the
system (\ref{eqsL24}) in this case is
\be
b=b_0t^{-2\nu},\quad
a=a_1\,t^{-2\nu}+a_2\,t^{-3\gamma\,\nu}-\frac{b_0}{3}\,t^{6(\gamma-1)\,\nu}
,\qquad \gamma\ne 2/3,\qquad \nu\equiv 1/(3\gamma-2)\label{eqsab}\ee
where $b_0,\, a_1,\,a_2$ are arbitrary integration constants. Since we are
assuming $\mu$  characterizes a material fluid (baryons plus photons and
possibly CDM), we have that $1\leq
\gamma\leq 2$ and hence $\nu$ in (\ref{eqsab}) is always positive.
Inserting (\ref{eqsab}) into (\ref{metric}), (\ref{eqmu}), (\ref{eqq}), and
(\ref{eqsL24}) leads to
\be \frac{1}{\Phi}=\frac{1}{a+br^2}=\frac{t^{3\gamma\nu}}{(-b_0/3)t^3+
(a_1+b_0r^2)t+a_2}\label{cf}
\ee
\be
8\pi\,\mu=\frac{12\left[(2\gamma-5/3)b_0t^3+(a_1+b_0r^2)\,t+(3\gamma-1)\,a_2
\right]\left[(-b_0/3)\,t^3+(a_1+b_0r^2)\,t+a_2\right]}{\nu^2\,t^{4(3\gamma-1)
\nu}}\label{eqmu_new}\ee
\be 8\pi\,q_r=\frac{2}{\nu}\,r\,t^{-3\gamma\nu},\qquad
8\pi\,q=\frac{8|b_0|
\left|(-b_0/3)\,t^3+(a_1+b_0r^2)\,t+a_2\right|\,\nu
r}{t^{6\gamma\nu}}\label{eqq_new}\ee
\be 8\pi\,\Lambda=\frac{(-8/3)b_0^2\,t^6+12a_1b_0\,t^4+16a_2K\,t^3+3a_2^2}
{t^{4(3\gamma-1)\nu}}\label{eqLL_new}\ee

Because of the apparent (and coordinate dependent) resemblance of
(\ref{metric}) to a spatially flat FLRW, it is tempting to assume that these
two metrics have common geometric features. For example, it is evident from
(\ref{eqmu_new}), (\ref{eqq_new}) and (\ref{eqLL_new}) that $\mu,\,\Lambda$ and
$q$ diverge as $t\to 0$ for $\nu>0$ and $1\leq
\gamma\leq 2$, hence we
can identify $t=0$ as the locus of a ``big bang'' singularity, analogous to the
FLRW big bang. In spatially flat FLRW spacetimes it is always possible to
assume that the coordinate range is given by $0<t<\infty$ and $0\leq r<\infty$,
so that $t\to\infty$ and $r\to\infty$ mark asymptotic future infinities in the
timelike, null and spacelike directions. However, for the models under
consideration the coordinate domain is necessarily restricted by the extra
condition that the conformal factor $1/\Phi$ be a bounded function. Also,
proper time along the worldlines of comoving observers is $\tau=\int{dt/\Phi}$
evaluated for fixed $(r,\theta,\phi)$, and so a sufficient condition for having
$\tau\to\infty$ occurs if the conformal factor $1/\Phi$ diverges, even if it
does so for finite values of the coordinates $t,r$.  From (\ref{cf}), this
occurs for all $a_1,\,a_2,\,b_0,\gamma$, since the equation $(-b_0/3)t^3+
(a_1+b_0r^2)t+a_2=0$ always has real roots in the coordinate domain $t>0,r\geq
0$,  defining the hypersurface
\be B=[t,r(t),\theta,\phi],\qquad r(t)=\left[\frac{(1/3)b_0t^3-a_1t-a_2}
{b_0t}\right]^{1/2}\label{eqB}\ee
which can be represented as a parametric curve in the $t,r$ plane. If one or
both of $a_1,\,a_2$ is zero, the boundary $B$ persists, though its
parametrization in the $t,r$ plane is simpler than (\ref{eqB}). The only
exception is if $b_0=0$, whence the solutions trivially reduce to FLRW.
Therefore, this feature is inherent to the models characterized by
(\ref{eqst}), (\ref{metric}) and (\ref{source}).

The fact that $\Phi^{-1}$ and $Y=r\Phi^{-1}$ diverge at (\ref{eqB}) means that
$B$ marks a spacetime boundary beyond which the spacetime manifold cannot be
extended.  An asymptotic past/future is then defined as the coordinate values
marked by $B$ which are reached by causal curves, either comoving observers
($r=$ const.) or radial null geodesics: ($v=t+r,\,w=t-r$).  From (\ref{cf}), it
is straightforward to prove that $\tau\to\infty$ holds as $B$ is reached by
future and past directed worldlines of comoving obervers. Also the affine
parameter of radial null geodesics diverge at spacetime points marked by $B$.
The coordinate domain of definition is then restricted by $|\Phi|>0$ and
depends on the signs of the constants $a_1,\,a_2,\,b_0$, specifying the form of
$B$ in the plane $t,r$.
>From the various  numerical values for these constants, we
eliminate all those cases in which the evolution of the comoving observers
occurs between two branches of $B$. The remaining cases display the two types
of evolution classified below:

\smallskip
\noindent
\underline{Case (i)}. If  $a_1,\,a_2,\,b_0$ are negative, $B$ lies in the
infinite past of all observers evolving towards their infinite future as
$t\to\infty$, and we have a null infinity analogous to that of a FLRW cosmology
(the infinite past is then marked by $B$). Using null coordinates the
asymptotic limit along outgoing radial null geodesics $v\to\infty$  is given by
\be
8\pi\,\mu\to\frac{16b_0^2(3\gamma-1)}{9\nu}\,v^{2(3\gamma-4)\nu},\qquad
8\pi\,q\to \frac{16b_0^2\nu}{3}\,v^{2(3\gamma-4)\nu},\qquad \Lambda\to
-\frac{8b_0^2}{3}\,v^{2(3\gamma-4)\nu}\label{asympt1}\ee
so that a regular null infinity requires $\gamma>4/3$  (otherwise, the affine
parameter has a finite limit as $t\to\infty$ and this locus marks a null
singularity). For a heat conducting shear-free fluid the weak energy condition
requires: $\rho+p=\gamma\mu>2q$ \cite{EC}, a relation that is satisfied by the
asymptotic forms (\ref{asympt1}) only for $2/3<\gamma<1$. Hence, this case is
unphysical.

\smallskip
\noindent
\underline{Case (ii)} If $a_1,\,a_2,\,b_0$ are positive, then
all worldlines of comoving observers start their evolution at $t=0$ (big bang)
and evolve towards their infinite future at $B$. From (\ref{eqmu_new}),
(\ref{eqq_new}) and (\ref{eqLL_new}), we have
\be
\frac{q}{\mu}=\frac{2\nu^3|b_0|t^2r}{(2\gamma-5/3)\,b_0t^3+(a_1+b_0r^2)\,t+(3\gamma-1)\,a_2}
\ee

\be\frac{\Lambda}{\mu}=\frac{\nu^2\left[\left(-(8/3)\,b_0t^6+12\,a_1t^4+16\,a_2t
^3\right)\,b_0 +3a_2^2
\right]}{12\left[(2\gamma-5/3)\,b_0t^3+(a_1+b_0r^2)\,t+(3\gamma-1)\,a_2
\right]\left[(-b_0/3)\,t^3+(a_1+b_0r^2)\,t+a_2\right]}\ee
so that near $t=0$ we obtain
\be\frac{q}{\mu}\to\frac{2\nu^3r}{(2\gamma-5/3)t}\label{qmu_bb}, \quad
 \frac{\Lambda}{\mu}\to\frac{\nu^2}{4(3\gamma-1)}\label{Lmu_bb}\ee
while at the boundary $B$ we have
\be
\mu\to 0,\qquad q\to 0,\qquad \rho\to \Lambda,\qquad p\to-\Lambda,\qquad
\Lambda\to\Lambda_{_B}=\Lambda(t_{_B})\ee
\be\left[\frac{q}{\mu}\right]_B
=\frac{3\nu^4\sqrt{|b_0|}\,t_{_B}^2r_{_B}}{2b_0t_{_B}^3+a_2}\label{qmu_B},
\quad
\Lambda_{_B}=\frac{4b_0\,t_{_B}^2[(a_1-2b_0r_{_B}^2)t_{_B}+2a_2]}{t_{_B}^{4(3
\gamma-1)\nu}}
\label{Lmu_B}\ee
where $t_{_B},\,r_{_B}$ are related by (\ref{eqB}). The limits (\ref{qmu_bb})
 indicate (for $1\leq\gamma\leq 2$) that the models are
matter dominated at the big bang ($0<\Lambda\ll\mu$), evolving towards a
$\Lambda$ dominated future at $B$ (represented by (\ref{qmu_B}).
 Notice that the asymptotic future state at $B$ can be
de Sitter  (or anti de Sitter)  and is not an asymptotically homogeneous state
since $\Lambda_{_B}$ depends on position (for each observer a different
constant value). For $r_{_B}=0$ (in (\ref{eqB})) we have that $\Lambda_{_B}>0$.
As $r_{_B}$ grows along $B$, $\Lambda_{_B}<0$, which implies that the total
energy density $\rho=\mu+\Lambda$ becomes negative asymptotically. Hence, for
the physical reasons, we exclude coordinate values $r>\bar r_{_B}$, where $\bar
r_{_B}$ satisfies $\Lambda_{_B}(\bar r_{_B})=0$.

The behavior of $q$ is compatible with the energy conditions, since $q\approx
\mu r/t \ll \mu$
holds all along the evolution, near the big bang and near $B$. It is still
necessary to find an adequate physical interpretation for this term, whether as
a heat flux or as a kinetic term associated with a velocity field or the dipole
of a kinetic theory distribution \cite{elliselst}. However, since we are
interested mainly in the asymptotic stage near $B$, as long as $r$ is
sufficiently small we will have $q\ll\mu $ and could consider $q$ as a residual
term.

Another feature of the models, absent in FLRW spacetimes, is the fact that
$\mu$ and $q$ diverge as $r\to\infty$ along hypersurfaces of constant $t$ that
do not intersect $B$, marking a point singularity at spacelike infinity. This
feature is also present for perfect fluid sources of (\ref{metric}),  see
\cite{cfpf}. Note that models similar to those examined here were considered
recently \cite{CJP}; however, the existence of the boundary (\ref{eqB}) was not
considered in the asymptotic study of those models.

The models discussed in this section illustrate how even simple inhomogeneous
spacetimes have a much richer geometrical structure that heavily constrains
their physical applicability. Pending a reasonable physical interpretation for
$q_a$  and provided we exclude sufficiently large values of $r$,   these
solutions are  inhomogeneous models that comply with the EGS criterion and
describe  a $\Lambda$ dominated scenario usually associated with the
``quintessence'' field \cite{RaPeeb,Wetterich,zlatov}.

\section{Discussion}

In this paper we have proven the existence and examined the physical viability
of a number of spacetimes which have been constructed to allow an isotropic
radiation field. Since these inhomogeneous spacetimes   satisfy the Copernican
principle (as far as the CMB is concerned), the question of finding methods of
testing the cosmological principle, and thus observationally testing whether
the universe is in fact an FLRW model, arises.

More precisely, it has been shown here and elsewhere~\cite{cla-bar99,BC} that
inhomogeneous universe models with non-geodesic observers obey the EGS
criterion. That is, inhomogeneous spacetimes have been found which allow every
observer to see an isotropic CMB. It has also been shown that a significant
subset of these models are consistent with other observational constraints, and
hence these models are consistent with observations even when the Copernican
principle is taken into account~-- and yet the models are not spatially
homogeneous~\cite{BC}. A potential problem with these particular models is that
the non-geodesic motion of the observers prohibits a barotropic equation of
state for perfect fluid matter. However, we have shown here that   more general
and physically viable cosmological models (with realistic matter)   allow an
isotropic radiation field. In particular, irrotational multi-fluid spacetimes
have been constructed which satisfy~(\ref{irf}). These cosmologies include two
non-comoving perfect fluids, which can be interacting or non-interacting, and
may or may not admit barotropic equations of state for the fluids. The fluids
are chosen to be non-comoving to allow for energy flux and anisotropic
pressures in the energy momentum tensor (otherwise the models may be written as
a single perfect fluid and correspond to the models studied
in~\cite{cla-bar99,BC}). Even in the acceleration-free case there are examples
of spacetimes that are not FLRW \cite{CM}. Similarly, models with a perfect
fluid and scalar field can be constructed in which the scalar field can have a
spatial gradient relative to the perfect fluid, which is taken to be comoving
with the isotropic radiation field. As a particular example, a class of
shear-free spherically symmetric, inhomogeneous  (quintessential) cosmologies
whose source is a heat conducting fluid and a scalar field were considered in
detail.

One of our key assumptions has been zero rotation. It has been shown that
rotating spacetimes which allow an isotropic radiation field may also be
constructed~\cite{ob1,ob2,obukov,carn}.

Other recent work also supports these conclusions. In the fundamental EGS
theorem \cite{EGS}, and here, it is assumed that all fundamental observers
measure the CMB temperature to be exactly isotropic during a time interval $I$
(defined by $t_E
\leq t \leq t_0$, where $t_E$ is the time of last scattering and $t_0$ is the
time of observation). Under this assumption the theorem then asserts that the
universe is {\em exactly} an FLRW model during this time interval. However, the
EGS theorem cannot be used to conclude that the physical universe is close to
an FLRW model since the CMB temperature can only be observed at one instant of
time on a cosmological scale. Hence it is of interest to ask what restrictions,
if any, can be placed on the anisotropy in the rate of expansion, assuming that
 all fundamental observers measure the CMB temperature to be
exactly isotropic at {\em some instant of time} $t_0$ only. On the basis of
continuity, it can then be argued that all fundamental observers will measure
the CMB temperature to be almost isotropic in some time interval of time of
length $\delta$ centered on $t_0$. This time interval could, however, be much
shorter than the time interval $I$. However, in \cite{LNW} it was shown that
{\em for a given time $t_0$}, there is a class of locally rotationally
symmetric  non-tilted dust  Bianchi type VIII spatially homogeneous
cosmological models such that at $t_0$ the CMB temperature is measured to be
isotropic by all fundamental observers, even though the overall expansion of
the universe is highly anisotropic at $t_0$.

In addition, the EGS theorem is of course not directly applicable to the real
universe since the CMB temperature is not exactly isotropic. This result has
consequently been generalized by \cite{almost} to the ``almost EGS theorem'',
which states that if all fundamental observers measure the CMB temperature to
be almost isotropic during some time interval in an expanding universe, then
the universe is described by an almost FLRW model during this time interval.
The dimensionless {\it shear parameter} and  the {\it Weyl parameter} were
introduced in \cite{WE}. Since the  Weyl curvature tensor is related to time
derivatives of the shear tensor, restricting the shear parameter to be small
does {\it not} guarantee that the Weyl parameter is small. Therefore a
necessary condition for the universe to be close to an FLRW model is that both
of the shear and Weyl parameters must be small. In the almost-EGS theorems the
dimensionless time and spatial derivatives of the
  multipoles are assumed to bounded by the multipoles themselves
  \cite{almost}. If this assumption is not satisfied,
then the CMB temperature observations {\it do not impose upper bounds on the
shear and Weyl parameters}, and hence do not establish that the universe is
close to FLRW. In \cite{NWL} a class of spatially homogeneous non-tilted
Bianchi type VII$_0$ dust models in which the CMB is treated as a test field or
a non-interacting radiation fluid was studied. To obtain the present CMB
temperature pattern, the photon energies were integrated numerically along the
null geodesics that connect points of emission on the surface of last
scattering with the event of observation at the present time. Wainwright et al.
\cite{WHU} then showed that the shear parameter tends to zero but the Weyl
parameter does not tend to zero at late times in these models.  In other words,
{\it although the models isotropize as regards the shear, the Weyl curvature
remains dynamically significant}. A variety of numerical simulations to
calculate anisotropy patterns of the CMB temperature in Bianchi VII$_0$ models
were explicitly performed \cite{NWL} to demonstrate that there exists
cosmological models that are not close to any FLRW model even though the
temperature of the CMB is almost isotropic in the sense that the observational
bounds on the quadrupole and octupole are satisfied.

It is clear, then, that the Copernican principle when combined with our
observations of the CMB does not imply the cosmological principle: that the
universe is homogeneous and isotropic. Recent work on this suggests that the
assumptions in the EGS (dust observers) and almost-EGS (small gradients of CMB
multipoles and dust observers) theorems are crucial to the conclusions;
weakening of any of these assumptions appears to negate the theorems almost
entirely. It is therefore important to ask what observations we need to test
the cosmological principle?

Assuming for the moment that the assumptions of the almost EGS theorem actually
hold in our universe, then one method by which the cosmological principle may
be tested is as follows: if we can observe the CMB as seen by some other
observers, then we can immediately confirm or reject the cosmological
principle. That is, if we find the CMB is as isotropic around these other
observers as we see it around us, we may conclude that we live in a homogeneous
universe. There is, in fact, a physical method by which we \emph{can} observe
the CMB as seen by other observers. It consists of light from the CMB being
scattered by hot gas in galaxy clusters in such a way as to allow us to observe
the anisotropy of the CMB as seen by that particular galaxy. This is known as
the Sunyaev-Zel'dovich (SZ) effect \cite{sun-zel69}, and has been suggested by
a number of authors as a possible means to test the Copernican and cosmological
principles \cite{goo95}.

This will only work of course if the initial assumptions of the EGS theorem
apply in our actual Universe. The original EGS theorem relies on the observers
in the universe being well described by a dust fluid~-- i.e., they are
geodesic. Indeed, in the almost EGS theorem, it was necessary to demand dust
observers to first order~-- i.e., more general matter was only allowed at
second order.

Any possible attempt to verify the cosmological principle by using methods such
as the SZ effect above, or any other method which relies on making observations
of the CMB from other locations, will fail. If we were living in a universe
found here or in \cite{cla-bar99}, for example, we would see exactly the same
effect: all observations of the CMB around other observers would be as
isotropic as the standard homogeneous FLRW models. It follows, therefore, that
\emph{the high isotropy of the CMB can never be used, on its own, to show our
universe is nearly homogeneous}.

On the other hand, we \emph{can} test the cosmological principle using the SZ
effect and methods like it \emph{provided} we can show definitively that our
universe is made of a dust-like fluid and that we travel on geodesics. Recent
observations suggesting quintessential matter making up a significant part of
the energy density of the universe throws this standard assumption into
question: there is no \emph{a priori} reason why this matter~-- whatever it
turns out to be~-- should be homogeneous (many dark matter theories allow
equations of state which are not dust also).

The geodesic assumption can be tested to some degree by local observations. If
we look closely at the recession velocity of galaxies close to us, then we can
detect a dipole moment in the relative velocities of these galaxies. This
deviation from the linear Hubble law is usually attributed to our local random
motion with respect to the local expansion rate of the universe, caused by our
gravitational infall into the Great Attractor. Acceleration will also leave its
mark on the linear Hubble law also as a dipole distribution in the direction of
the CMB dipole, but as a dipole which grows linearly with
distance~\cite{cla-bar99}. With sufficiently accurate knowledge of the local
distribution of galaxies, the two effects can be disentangled as our bulk
gravitational motion affects the Hubble law irrespective of distance. Current
knowledge of our local group motion will only provide relatively weak
constrains on acceleration (roughly $\udot/H_0\sim0.1$; compare this with
$\sigma/H_0^2\sim 10^{-5}$ from almost-EGS CMB observations~\cite{MES5}).

If we assume, for arguments sake, that the results of such a study reveal that
the local dipole is entirely accounted for by our peculiar velocity, then what
can we say about the spatial homogeneity of our universe? If the acceleration
around our location is small, then we may assert the Copernican principle and
ascertain that all observers in the universe follow geodesics; therefore we may
apply the EGS theorem (assuming the SZ measurements measure small enough
anisotropies of the CMB around other observers) and deduce that the
cosmological principle is a valid assumption. This would yield tremendous
support for our faith in the standard model.

Of course, acceleration is just one possible inhomogeneity which causes
problems: others are rotation, Weyl curvature, and anisotropy of the energy
momentum tensor. These all leave their mark to varying degrees to the
anisotropy of the magnitude-redshift and number-count-redshift relations, but
at higher order in redshift (if expanded in a power series in redshift) than
acceleration, so are much harder to detect.

What should be clear from this argument is that the cosmological principle is
simply untestable. Even if SZ measurements reveal the Copernican principle to
be true, accurate determination of anisotropies of the magnitude-redshift
relation at high redshift are out of the question for the foreseeable future.

\subsection{Acknowledgements}
We would like to that Martin Hendry, Stephane Rauzy and Kenton D'Mellow for
useful discussions. RAS would like to thank Dalhousie University for
hospitality while this work was carried out. The work was supported, in part,
by NSERC.

\subsection*{A. Appendix. The energy momentum tensor of conformally
related spacetimes}
\label{conformal trans section}

A  conformal transformation  is an angle preserving transformation that changes
lengths and volumes. The importance of these types of transformations lies in
the fact that, under a conformal transformation, \emph{the  causal structure of
the spacetime is preserved.} The Weyl tensor, $C_{abc}^{\phantom{abc}d}$, is
invariant, so that a conformal transformation will introduce no tidal forces or
gravitational waves; that is, a conformal transformation will only introduce
`non-gravitational' forces and matter into the new spacetime (by changing
$R_{ab}$ and thus the matter tensor $T_{ab}$ via Einstein's equations).

We will discuss conformal transformations and their 1+3 splitting here as it is
a useful tool for constructing new spacetimes from old, especially if one is
after spacetimes with an isotropic radiation field. The conformal
transformation is defined by
\be
g_{ab}=e^{2Q}\ghat_{ab},\pha u^a=e^{-Q}\uhat^a,\pha u_a=e^Q\uhat_a;
\ee
where $Q>0$ is an arbitrary function, $u^a$ is a velocity vector with respect
to $g_{ab}$: $g_{ab}u^au^b=u_au^a=-1$; and $\uhat^a$ is the conformally related
(parallel) velocity vector, and is normalised with respect to $\ghat_{ab}$:
$\ghat_{ab}\uhat^a\uhat^b=-1$. The covariant derivative of any one-form field
$v_a$ transforms as
\be
\del_av_b=\delhat_av_b-2Q_{(a}v_{b)}+g_{ab}Q^cv_c,
\ee
where $Q_a\equiv Q_{,a}=\sdel_aQ-\dot{Q}u_a$. The expansion ($\theta=\div u$),
acceleration ($\udot^a=u^b\del_bu^a$), rotation ($\omega_{a}=-\:12\curl u_a$),
and shear ($\sigma_{ab}=\sdel_{\< a}u_{b\rangle}$) of the two velocity
congruences are related by:
\ba
\thhat\li=\li e^{Q}(\theta-3\dot{Q})\nonumber\\
\ahat_a\li=\li\dot{u}_a-\sdel_a Q\nonumber\\
\rothat_a\li=\li\omega_a\nonumber\\
\hat{\sigma}_{ab}\li=\li e^{-Q}\sigma_{ab}.\label{kinematic_trans}
\ea
The equation for the acceleration corrects equation~(6.14) of \cite{Kramer}.
These show that a conformal transformation may induce acceleration and
expansion into the new spacetime, but not shear or rotation: in particular, a
conformally flat model must have vanishing shear and rotation
 \cite{cla-bar99,BC}.  With respect to
$g_{ab}$, a dot denotes time differentiation along the fluid flow,
 $\dot{F}_{ab}=u^c\del_c F_{ab}$, and $\sdel_a$ is the spatial derivative
projected orthogonal to the flow lines,   $\sdel_c F_{ab}=h^d_{\pha c}
h^e_{\pha a} h^f_{\pha b} \del_d F_{ef}$, where $h_{ab}=g_{ab}+u_au_b$ is the
usual projection tensor.

The Einstein tensor transforms as
\be
G_{ab}=\Ghat_{ab}-2\del_aQ_b-2Q_aQ_b+g_{ab}\left[2\del_cQ^c-Q^2\right],
\label{G_transform}
\ee
where $\Ghat_{ab}$ is the Einstein tensor of $\ghat_{ab}$, and $Q^2=Q_aQ^a$.
For clarity, we decompose derivatives of $Q$ into time and space derivatives:
\ba
&&Q_a=\sdel_a Q-\qdot u_a,\nonumber\\
\del_bQ_a&=&u_au_b\left(\ddot{Q}-\udot^c\sdel_cQ\right)+2u_{(a}\left[-\sdel_{b)}
\qdot+
\frac{1}{3}\theta\sdel_{b)}Q+\left(\sigma_{b)c}+\eta_{b)dc}\omega^d
\right)
\sdel^cQ\right]\nonumber\\&&+\:13h_{ab}\left[\sdel^c\sdel_cQ-\dot{Q}\theta\right]
- \dot{Q}\sigma_{ab}+\sdel_a\sdel_bQ.
\label{Q_derivatives}
\ea
We also write $\hat T_{ab}=\Ghat_{ab}$, and $T_{ab}=G_{ab}$ as general fluids,
 both with respect to $u^a$:
\ba
&\hat{G}_{ab}=\hat\mu\hat u_a\hat u_b +\hat{p}\hat h_{ab} +2\hat{q}_{(a}\hat
u_{b)}+\hat{\pi}_{ab},\label{Gbar}
\\ &{G}_{ab}=\mu u_a u_b +p
h_{ab}+2{q}_{(a}u_{b)}+{\pi}_{ab};\label{G_fluid}
\ea
where $\{\hat\mu,\hat{p},\hat{q}_a, \hat{\pi}_{ab}\}$, and
$\{\mu,p,q_a,\pi_{ab}\}$ are the energy density, isotropic pressure, heat flux,
and anisotropic pressure of $\Ghat_{ab}$ and $G_{ab}$ respectively.

We can decompose $G_{ab}$ given by~(\ref{G_transform}) into the fluid variables
in~(\ref{G_fluid}) by using~(\ref{Q_derivatives}) in the following covariant
manner:
\ba
\mu\li=&
u^au^bG_{ab}=e^{2Q}\hat\mu-3\qdot\left(\qdot-\frac{2}{3}\theta\right)-
2\sdel_a\sdel^aQ+\sdel_aQ\sdel^aQ,\\
p&=&\frac{1}{3}h^{ab}G_{ab}=e^{2Q}\hat{p}+\left(\qdot-\frac{4}{3}\theta\right)
\qdot-2\ddot{Q}+
\frac{4}{3}\sdel_a\sdel^aQ-\frac{5}{3}\sdel_aQ\sdel^aQ
+2\udot_c\sdel^cQ,\nonumber
\\ q_a&=&-u^bG_{\langle
a\rangle b}=e^{Q}\hat{q}_a\forget{-2\sdel_a\qdot+2\left(\frac{1}{3}\theta-
\qdot\right)
\sdel_aQ+2\left(\sigma_a^{\pha b}+\omega_a^{\pha b}\right)\sdel_bQ}
+2\dot Q\left(\sdel_aQ-\udot_a\right)+2h_a^{~b}\left(\sdel_bQ\right)^{.}
+4\eta_{abc}\omega^b\sdel^cQ
\\
\pi_{ab}&=& G_{\langle ab\rangle}=\hat{\pi}_{ab}+2\qdot\sigma_{ab}-2
\sdel_{\< a}\sdel_{b\>}Q-2\sdel_{\< a}Q\sdel_{b\>}Q.
\ea
In the energy flux equation we used the identity
\be
\sdel_a\dot Q=h_a^{~b}\left(\sdel_bQ\right)^{\cdot}-\dot Q\udot_a
+\:13\theta\sdel_aQ +\sigma_{a}^{~b}\sdel_bQ+ \eta_{abc}\omega^b\sdel^cQ .
\ee

\end{document}